\documentclass[fleqn,usenatbib,useAMS]{mnras}

\usepackage{newtxtext,newtxmath}

\usepackage[T1]{fontenc}

\DeclareRobustCommand{\VAN}[3]{#2}
\let\VANthebibliography\thebibliography
\def\thebibliography{\DeclareRobustCommand{\VAN}[3]{##3}\VANthebibliography}

\usepackage{graphicx}	
\usepackage{amsmath}	
\usepackage{orcidlink}

\newcommand{\msun}{\,{\rm M}_{\sun}}

\title[Oldest magnetic white dwarfs]{Younger age for the oldest magnetic white dwarfs}

\author[S. Ginzburg]{
Sivan Ginzburg$^{\orcidlink{0000-0002-3751-4553}}$\thanks{E-mail: \href{mailto:sivan.ginzburg@mail.huji.ac.il}{sivan.ginzburg@mail.huji.ac.il}}
\\
Racah Institute of Physics, The Hebrew University, Jerusalem 9190401, Israel
}

\date{Accepted XXX. Received YYY; in original form ZZZ}

\pubyear{\the\year{}}

\begin{document}
\label{firstpage}
\pagerange{\pageref{firstpage}--\pageref{lastpage}}
\maketitle

\begin{abstract}
Sufficiently old white dwarfs cool down through a convective envelope that directly couples their degenerate cores to the surface. Magnetic fields may inhibit this convection by stiffening the criterion for convective instability. We consistently implemented the modified criterion in the stellar evolution code \textsc{mesa}, and computed the cooling of white dwarfs as a function of their mass and magnetic field $B$. In contrast to previous estimates, we find that magnetic fields can significantly change the cooling time $t$ even if they are relatively weak $B^2\ll 8\upi P$, where $P$ is the pressure at the edge of the degenerate core. Fields $B\gtrsim 1\textrm{ MG}$ open a radiative window that decouples the core from the convective envelope, effectively lowering the luminosity to that of a fully radiative white dwarf. We identified a population of observed white dwarfs that are younger by $\Delta t\sim$ Gyr than currently thought due to this magnetic inhibition of convective energy transfer -- comparable to the cooling delay due to carbon--oxygen phase separation. In volume-limited samples, the frequency and strength of magnetic fields increase with age. Accounting for magnetic inhibition is therefore essential for accurate cooling models for cosmic chronology and for determining the origin of the magnetic fields. 
\end{abstract}




\section{Introduction}\label{sec:intro}

White dwarfs cool down over billions of years by transporting heat outwards from their interior reservoirs through their envelopes. In the simplest model \citep{Mestel1952}, the white dwarf is composed of a degenerate core, which is roughly isothermal thanks to the efficient heat conduction of degenerate electrons, surrounded by an ideal gas envelope that insulates the core and regulates its cooling by photon diffusion (i.e. radiation). Such cooling models relate the luminosity and effective temperature of a white dwarf to its mass and age (more precisely, the cooling time), enabling white dwarfs to serve as cosmic clocks and thus establish the age of various stellar populations -- `Galactic history is written in the white dwarf stars' \citep{Winget1987,HansenLiebert2003,WingetKepler2008}.  

However, an accurate age determination requires accurate cooling models, and the cooling of old (several Gyr) white dwarfs in particular involves several complications, such as the crystallization of their cores \citep{VanHorn1968}, and the reduction in their heat capacity as they transition to the Debye regime \citep{MestelRuderman1967}. See \cite{Fontaine2001,Althaus2010,Renedo2010}, for comprehensive reviews. Recently, accurate measurements by the {\it Gaia} satellite \citep{Gaia2018,Gaia2018n} emphasized the need for theoretical cooling models with sub-Gyr precision for the bulk of the white dwarf population \citep{Tremblay2019}, as well as additional physics (such as neon distillation) to explain the multi-Gyr cooling delays of a fraction of stars that lie on the `Q branch' \citep{Cheng2019,Bauer2020,Blouin2021,Camisassa2021,Bedard2024}. 

One of the important deviations from the simple cooling model of \citet{Mestel1952} is convection in the white dwarf's envelope. Sufficiently cold white dwarfs develop outer convection zones that gradually expand inwards as the stars continue to cool down. While convection can transport heat faster than radiation (for the same temperature gradient), it initially does not affect the white dwarf's cooling rate. The reason is that the white dwarf's interior degenerate core, which dominates the heat capacity, is decoupled from the outer convective envelope by an insulating radiative layer. This radiative bottleneck dictates the white dwarf's cooling until `convective coupling' occurs -- when the convection zone penetrates deep enough and reaches the degenerate core, at an age of a few Gyr \citep{Fontaine2001}.

Convection is sensitive to magnetic fields, and a large fraction of white dwarfs are strongly magnetized \citep[see][for reviews]{Ferrario2015,Ferrario2020}. The origin of white dwarf magnetism is still an open question. Possible explanations include a fossil field inherited from a previous stage of stellar evolution \citep{Angel1981,BraithwaiteSpruit2004,Tout2004,WickramasingheFerrario2005}, a dynamo operating during a double white dwarf merger \citep{GarciaBerro2012} or during a common envelope event \citep{RegosTout1995,Tout2008,Nordhaus2011}, and an internal dynamo driven by crystallization or distillation in the white dwarf's core \citep{Isern2017,Schreiber2021Nat,Ginzburg2022,Fuentes2023,Fuentes2024,Castro-Tapia2024,Lanza2024,MontgomeryDunlap2024}. In some of these mechanisms, magnetic fields emerge at the white dwarf's surface only after several Gyr of cooling and magnetic diffusion, such that accurate cooling models are required to assess the relative contribution of different magnetization channels to the overall population \citep{BagnuloLandstreet2022,BlatmanGinzburg2024a,BlatmanGinzburg2024b,Castro-Tapia2024_diff}.

\begin{figure*}
	\includegraphics[width=\textwidth]{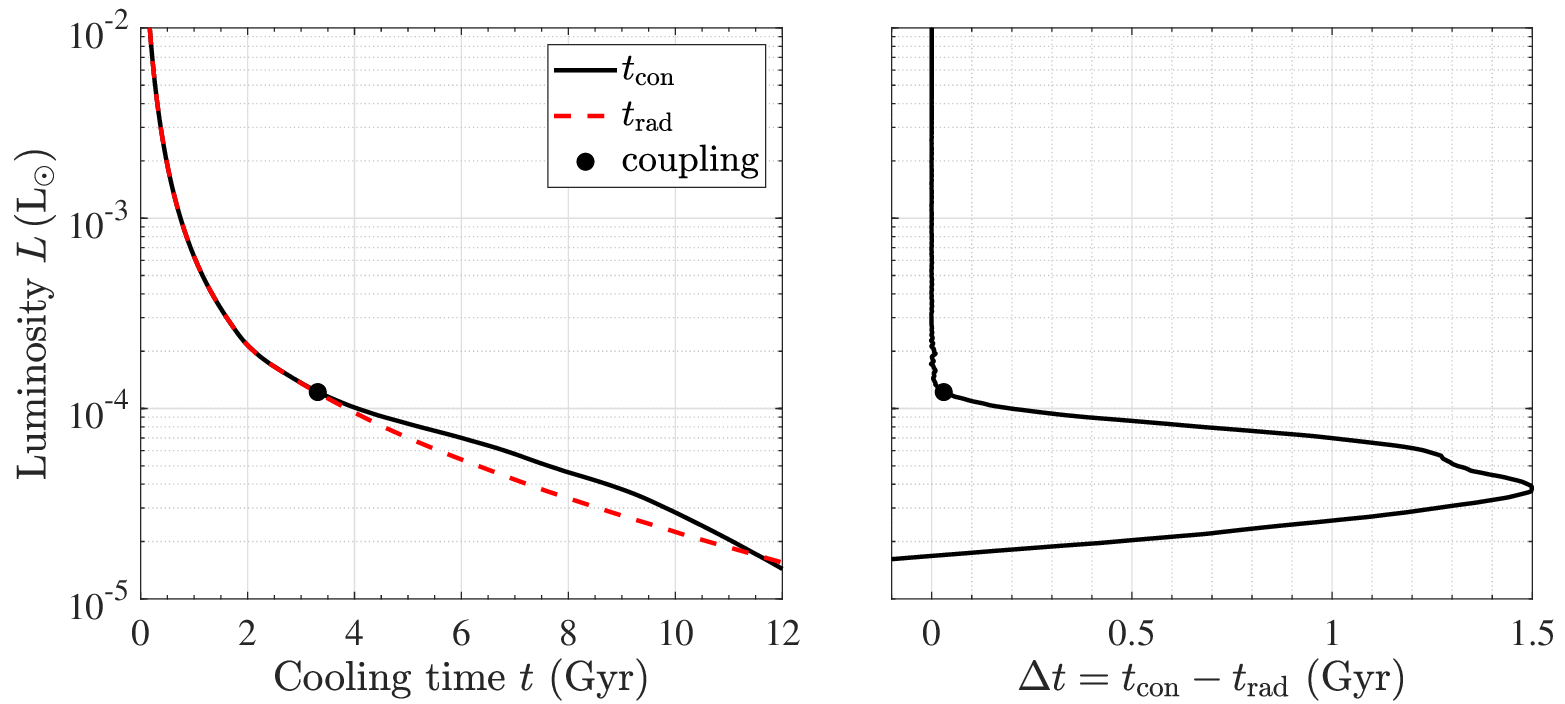}
    \caption{{\it Left panel:} Cooling tracks of $0.6\msun$ white dwarfs with (solid black line) and without (dashed red line) convection. Convection has no effect on the cooling rate until convective coupling occurs (circular marker) -- when the outer convective envelope reaches the conducting degenerate core -- eliminating the intermediate radiative bottleneck. From this moment, convection initially increases the luminosity at a given age, but eventually depletes the white dwarf's thermal energy reservoir faster, leading to lower luminosities compared to the fully radiative model. {\it Right panel:} The cooling delay, i.e. the difference in the cooling time required to reach a given luminosity between the convective and radiative models.}\label{fig:delay}
\end{figure*}

\citet{Valyavin2014} suggested that the inhibition of convection by strong magnetic fields may significantly alter white dwarf cooling rates. \citet{Tremblay2015} demonstrated, on the other hand, that due to the radiative bottleneck, any change to the efficiency of convection can have an impact only after convective coupling. Since almost all known magnetic white dwarfs at the time were younger than that, they concluded that magnetic fields do not affect the observed population.     

Here, motivated by new observations of cold magnetic white dwarfs over the past decade, we revisit the effects of magnetic fields on white dwarf cooling. \citet{Tremblay2015} computed cooling tracks with convection entirely shut off -- mimicking the influence of the strongest magnetic fields. We modify the criterion for convective instability consistently in order to evaluate the inhibition of convective energy transfer by much weaker fields as well.\footnote{Note that convective chemical mixing is not necessarily inhibited to the same extent as the energy transfer \citep[e.g.][]{Cunningham2021}.}

The remainder of this letter is organized as follows. In Section \ref{sec:times}, we repeat the \citet{Tremblay2015} exercise and compute the cooling time difference between regular and strongly magnetized white dwarfs as a function of mass and age. In Section \ref{sec:magnetic}, we modify the criterion for convective instability and compute cooling tracks as a function of the magnetic field strength. We relate our results to the observed white dwarf population in Section \ref{sec:obs}, and summarize our conclusions in Section \ref{sec:summary}.

\section{Cooling times}\label{sec:times}

In Fig. \ref{fig:delay}, we use the test suite \texttt{wd\_cool\_0.6M} of the stellar evolution code \textsc{mesa} \citep{Paxton2011,Paxton2013,Paxton2015,Paxton2018,Paxton2019,Jermyn2023}, version r23.05.1, to compute the cooling of a $0.6\msun$ carbon--oxygen white dwarf. The initial white dwarf model at the onset of this cooling phase was created using the test suite \texttt{make\_co\_wd}, which evolves a main sequence progenitor star through the various stages of stellar evolution. Similarly to \citet{Tremblay2015}, we also compute another cooling track, where convection is artificially disabled for the entire cooling phase -- mimicking inhibition by strong magnetic fields. 

As seen in Fig. \ref{fig:delay} and explained by \citet{Tremblay2015}, the white dwarf's cooling rate is sensitive to convection only at cooling times $t\gtrsim 3\textrm{ Gyr}$ -- after the convective envelope penetrates into the degenerate core. This convective coupling between the interior and the envelope is like opening the door of an oven: it initially increases the luminosity $L$ coming out from the hot core (compared to a radiative model), but the enhanced cooling rate depletes the core's thermal energy reservoir faster, such that in the long run ($t\gtrsim 11\textrm{ Gyr}$) convection actually reduces the luminosity \citep{Fontaine2001,Tremblay2015}. In this letter we focus on intermediate times, when convection manifests as a delay in the cooling process (i.e. convection keeps the luminosity $L$ higher for a longer time $t$). By comparing the right panel of our Fig. \ref{fig:delay} to fig. 5 of \citet{Bauer2023}, we see that the cooling delay due to convection is comparable (and even larger) than cooling delays due to other processes that are currently being studied, such as carbon--oxygen phase separation during crystallization, which delays the cooling at similar luminosities (for a $0.6\msun$ white dwarf). Magnetic fields that are strong enough to inhibit convection are therefore as important as other processes (which have received greater attention) for calculating accurate cooling tracks. 

\begin{figure}
\includegraphics[width=\columnwidth]{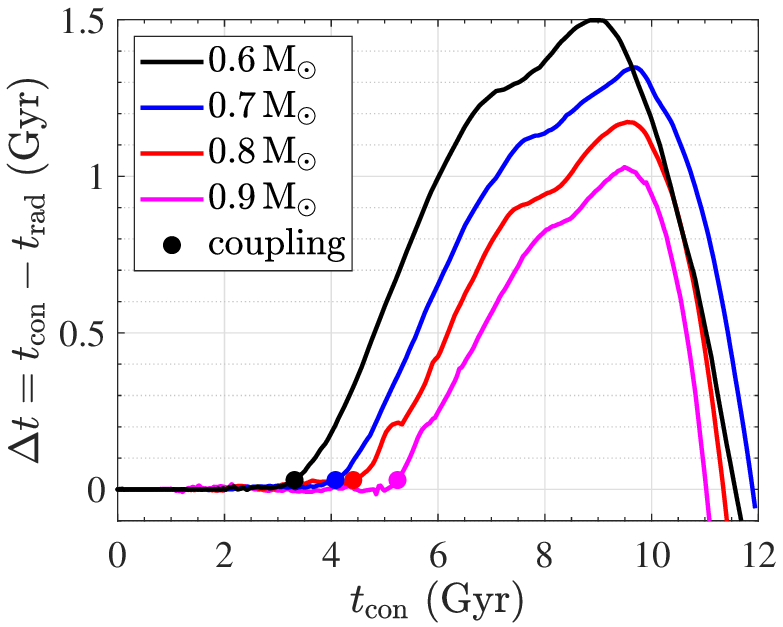}
\caption{Same as the right panel of Fig. \ref{fig:delay}, i.e. the cooling delay caused by convection, but displayed as a function of the nominal cooling time $t_{\rm con}$ (with convection enabled), and for several white dwarf masses. Circular markers indicate the moment of convective coupling.}
\label{fig:CoolChain}
\end{figure}

In Fig. \ref{fig:CoolChain} we repeat the calculation of the cooling delay caused by convection for white dwarfs with different masses, and also present the delay as a function of the nominal cooling time (i.e. with convection enabled) $t_{\rm con}$ instead of the luminosity. The initial white dwarf models are generated by evolving main sequence stars with different initial masses using the test suite \texttt{make\_co\_wd}, identically to \citet{BlatmanGinzburg2024a}. Our models have helium layers weighing $m_{\rm He}\sim 10^{-3}-10^{-2}\msun$ and relatively massive hydrogen atmospheres $m_{\rm H}\sim 10^{-5}-10^{-4}\msun$. The cooling rates of white dwarfs with helium-dominated atmospheres ($m_{\rm H}\sim 10^{-10}\msun$) differ significantly \citep[e.g.][]{BedardSpec2024}.
As seen in Fig. \ref{fig:CoolChain}, the age when convective coupling happens increases with mass. \citet{Fontaine2001} found similar convective coupling ages in their fig. 2, where they are also compared to the onset of crystallization. We see that convection is responsible for cooling delays of $\Delta t\sim{\rm Gyr}$, which peak at ages of $t\sim 6-10\textrm{ Gyr}$. If magnetic fields are strong enough to inhibit convection, magnetic white dwarfs at these ages may therefore be about 10 per cent younger than currently assumed. 

\section{Critical magnetic field}\label{sec:magnetic}

In Section \ref{sec:times} we established that convection significantly delays the cooling of old white dwarfs. The question is how strong does a magnetic field $B$ has to be in order to inhibit convection and shorten this delay. \citet{Tremblay2015} estimated a critical magnetic field $B\sim 10-100\textrm{ MG}$ by requiring that the magnetic energy density exceeds the pressure $P$ at the base of the convection zone $B^2\sim 8\upi P$, at the moment of convective coupling. 

\begin{figure}
\includegraphics[width=\columnwidth]{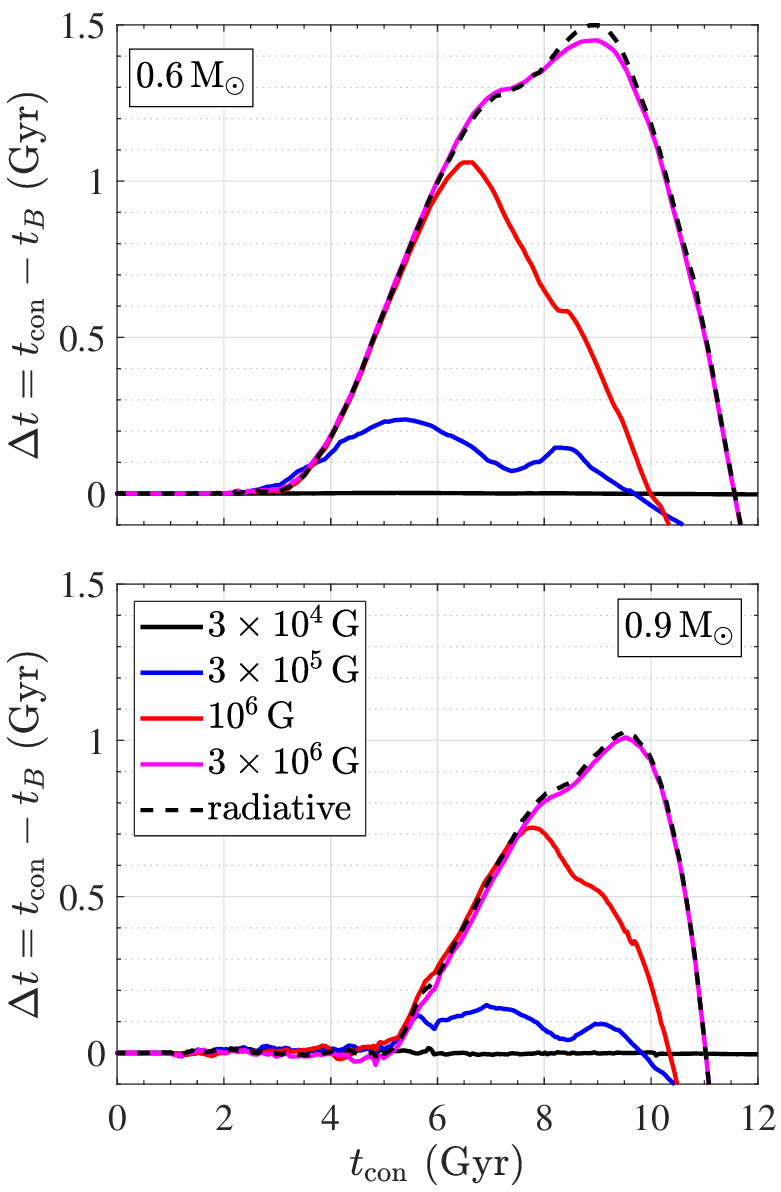}
\caption{The cooling time difference $\Delta t$ between non-magnetic white dwarfs (our nominal $t_{\rm con}$) and magnetic stars with a field strength $B$ that is given in the legend. Magnetic white dwarfs with $B\approx 3\textrm{ MG}$ cool down almost as if they were completely radiative (dashed black lines).}
\label{fig:MagneticPanels}
\end{figure}

We take a different approach, and consider the criterion for convective instability. Neglecting compositional gradients, convection develops when $\nabla_{\rm rad}>\nabla_{\rm ad}$, where $\nabla\equiv {\rm d}\,\ln T/{\rm d}\,\ln P$ denotes the temperature gradient (this is the Schwarzschild criterion). If the radiative gradient $\nabla_{\rm rad}$ required to carry the luminosity exceeds the adiabatic gradient $\nabla_{\rm ad}$ that fast convection maintains, then convective instability develops, such that the actual temperature gradient is given by $\nabla=\min(\nabla_{\rm rad},\nabla_{\rm ad})$. \citet{GoughTayler1966} extended this criterion to include a magnetic field, showing that convection develops when\footnote{The $4\upi$ factor is missing in some of the equations of \citet{GoughTayler1966} because they use rationalized units; see \citet{MullanMacDonald2001}.} 
\begin{equation}\label{eq:gough}
    \nabla_{\rm rad}>\nabla_{\rm ad}+\frac{B^2}{B^2+4\upi\gamma P},
\end{equation}
where $\gamma\equiv 1/(1-\nabla_{\rm ad})$ is the adiabatic index.
We implemented this criterion in \textsc{mesa} by using the \texttt{other\_mlt\_results} hook and setting the temperature gradient to 
\begin{equation}\label{eq:gough_mesa}
\nabla=\min\left(\nabla_{\rm rad},\nabla_{\rm ad}+\frac{B^2}{B^2+4\upi P}\right),    
\end{equation}
where we have omitted $\gamma\approx 5/3$ for simplicity, leading to a minor change in the critical $B$.
The results of our magnetic cooling tracks (compared to non-magnetic stars) are presented in Fig. \ref{fig:MagneticPanels}.

At face value, equation \eqref{eq:gough} seems equivalent to the criterion used by \citet{Tremblay2015}. When $B^2\gtrsim 8\upi P$, the stability criterion changes significantly -- disrupting convection (unless $\nabla_{\rm rad}>\nabla_{\rm ad}+1$). However, as seen in Fig. \ref{fig:MagneticPanels}, when incorporating the \citet{GoughTayler1966} criterion consistently into the stellar structure and cooling model, we find that even much weaker magnetic fields are sufficient to inhibit convective energy transfer. Cooling tracks with $B\approx 3\textrm{ MG}$ -- orders of magnitude weaker than assumed by \citet[][see their fig. 4]{Tremblay2015} -- are indistinguishable from fully radiative models (where convection is artificially disabled throughout the star). When the field is reduced further, to below $B\sim 10^5\textrm{ G}$, the cooling track converges to that of the nominal non-magnetic star.

\begin{figure}
\includegraphics[width=\columnwidth]{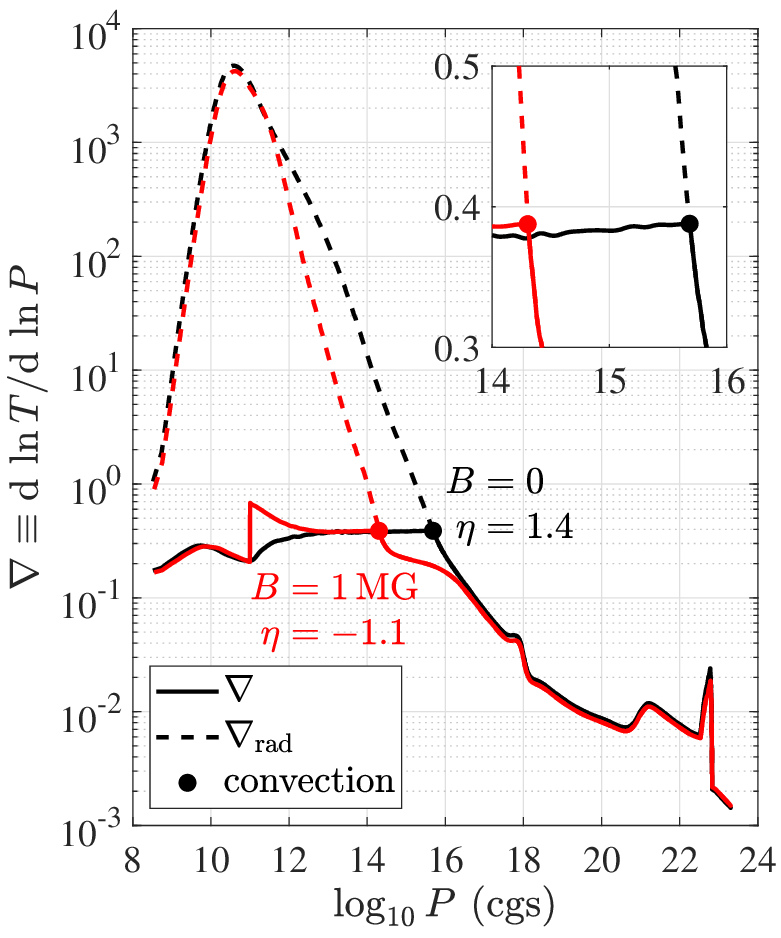}
\caption{The radiative (dashed lines) and actual (solid lines) temperature gradients inside a $0.6\msun$ white dwarf at $t=4\textrm{ Gyr}$ for $B=0$ (black) and $B=1\textrm{ MG}$ (red). The magnetic field $B$ is implemented by applying equation \eqref{eq:gough_mesa} for pressures $P>10^{11}\textrm{ dyn cm}^{-2}$ (for numerical convenience, see text). Convection operates to the left of the circular markers, for which we provide the value of the electron degeneracy parameter $\eta$. The inset zooms in on this boundary and demonstrates that even a small change ($\ll 1$) in $\nabla$ is enough to open a substantial radiative window that decouples the degenerate ($\eta>0$) core from the convective envelope, such that the white dwarf cools down as if it were completely radiative.}
\label{fig:prof}
\end{figure}

We can understand this sensitivity to weak magnetic fields by taking a closer look at the temperature gradients about a Gyr after convective coupling in Fig. \ref{fig:prof}. 
In the non-magnetic case, convection penetrates up to a pressure $P\sim 10^{16}\textrm{ dyn cm}^{-2}$, reaching the degenerate core (the electron degeneracy parameter $\eta>0$).\footnote{$\eta k T$ is the electron chemical potential, where $k$ is the Boltzmann constant, such that $e^\eta$ is related to the ratio of degeneracy to ideal pressures; see \citet{Kippenhahn2012}.} As dictated by equation \eqref{eq:gough_mesa} and demonstrated in Fig. \ref{fig:prof}, a relatively weak magnetic field $B\approx 1\textrm{ MG}$ raises the effective adiabatic gradient $\nabla_{\rm ad}\approx 0.4$ by a tiny amount because $B^2/(4\upi P)\ll\nabla_{\rm ad}\lesssim 1$ (much less than the \citealt{Tremblay2015} criterion). None the less, this small increase is sufficient to push the base of the convective envelope to a much lower pressure $P\sim 10^{14}\textrm{ dyn cm}^{-2}$, which is decoupled from the degenerate core ($\eta<0$). The radiative window that is thus opened between the convective envelope and the degenerate core acts as a bottleneck and decreases the luminosity (as can be seen by the lower $\nabla_{\rm rad}$, which pushes the convection further, self-amplifying the effect). In fact, as demonstrated in fig. 9 of \citet{Tremblay2015}, the optical depth of the white dwarf's envelope is dominated by the higher pressure at $\eta\approx 0$, such that once a radiative window that spans about a decade in pressure (around $\eta\approx 0$) is opened, the white dwarf effectively cools down as a fully radiative star (lower pressures almost do not contribute to the optical depth).   

We note that for numerical convenience we applied equation \eqref{eq:gough_mesa} -- which for $P\ll B^2$ adds $\approx 1$ to $\nabla_{\rm ad}$ -- only for $P>10^{11}\textrm{ dyn cm}^{-2}$. As seen in Fig. \ref{fig:prof}, this addition to $\nabla_{\rm ad}$ may open a second radiative window at very low pressures, requiring special attention. In this letter we focus on the vicinity of the degenerate boundary ($\eta\approx 0$), and defer the effects of magnetic fields on the shallower atmosphere to future work.  
None the less, as demonstrated in Figs \ref{fig:MagneticPanels} and \ref{fig:prof} for $B=1\textrm{ MG}$, even white dwarfs with unperturbed convection in their upper atmosphere cool down as if they were fully radiative, because the bottleneck for energy transfer is deeper inside the star, where $\eta\approx 0$. For this reason, our results are not sensitive to the treatment of convection at much lower pressures.

Although the \citet{GoughTayler1966} criterion is widely used in solar and stellar astrophysics, we are not aware that it has been tested against three-dimensional magnetohydrodynamic simulations. \citet{Tremblay2015}, on the other hand, compared their criterion to such simulations, albeit only for the upper atmosphere, much weaker magnetic fields $B\sim 10^3\textrm{ G}$, and higher effective temperatures $T_{\rm eff}\approx 10,000\textrm{ K}$. The original \citet{Tremblay2015} criterion is presented in the form of a sharp cutoff at $P=B^2/(8\upi)$. In reality, however, we might expect a smoother transition from strong to weak convection as a function of the pressure $P$. A reasonable smoothing of the \citet{Tremblay2015} criterion, or a consistent mixing-length theory, will yield an expression similar to equation \eqref{eq:gough} -- i.e. an addition $\sim B^2/P\ll 1$ to the effective $\nabla_{\rm ad}$ at high pressures \citep[see][]{BessilaMathis2024}. As explained above, this small deviation is responsible for decoupling the degenerate core from the convective envelope. We therefore presume that our result that relatively weak magnetic fields $B^2\ll 8\upi P$ are sufficient to change the cooling time is general and does not depend on the specific criterion used.

\section{Observations}\label{sec:obs}

\begin{figure}
\includegraphics[width=\columnwidth]{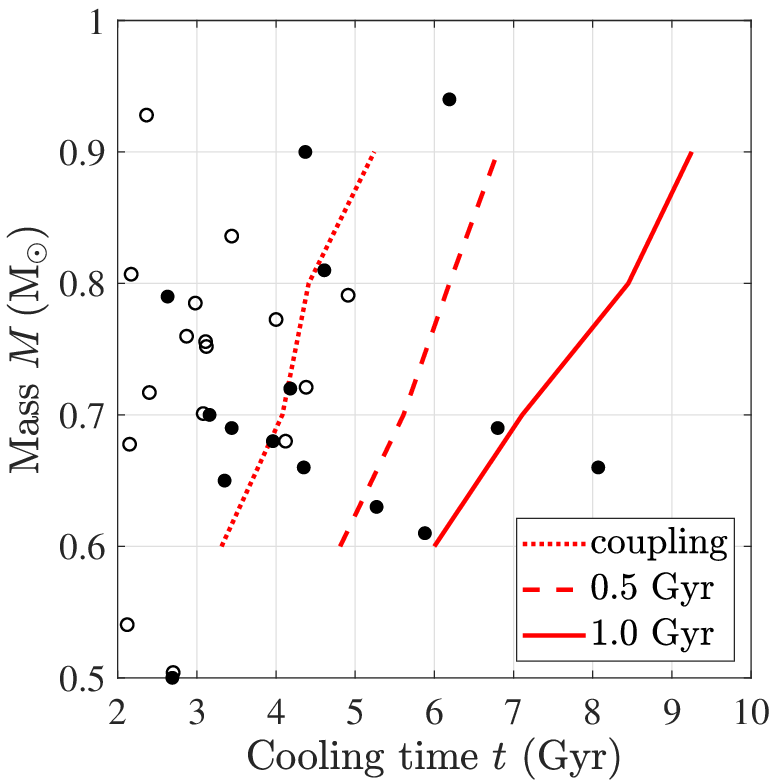}
\caption{Magnetic white dwarfs with $B> 1\textrm{ MG}$ from the 20 pc volume-limited sample of \citet[][filled circles]{BagnuloLandstreet2022}, and from the (mostly) magnitude-limited sample of \citet[][open circles]{Hardy2023}, which is biased towards shorter cooling times. The red lines are computed from Fig. \ref{fig:CoolChain}, and they indicate convective coupling (dotted), the first time $\Delta t=0.5\textrm{ Gyr}$ (dashed), and the first time $\Delta t=1.0\textrm{ Gyr}$ (solid). The figure demonstrates that several magnetic white dwarfs in our local 20 pc volume are younger than currently thought by $\sim 1\textrm{ Gyr}$ -- more than 10 per cent of their cooling age.}
\label{fig:obs}
\end{figure}

In Fig. \ref{fig:obs} we estimate how many white dwarfs are sufficiently old and magnetic for their cooling time to be affected by magnetic fields. We plot the observed white dwarfs with magnetic fields above the critical value of $B\sim 1\textrm{ MG}$ (see Section \ref{sec:magnetic}) from the volume-limited sample of \citet{BagnuloLandstreet2022} and from the \citet{Hardy2023} sample, which is derived from the Montreal White Dwarf Database\footnote{\url{https://www.montrealwhitedwarfdatabase.org}} \citep{Dufour2017}. The stars in the latter sample are mainly drawn from magnitude-limited surveys, over-representing short cooling times (there might be some overlap between the two samples, but this does not affect our conclusions). 

We overlay the observations with $t(M)$ lines indicating the moment of convective coupling, and the age when the cooling time difference $\Delta t$ first exceeds 0.5 and 1.0 Gyr, as a function of the white dwarf's mass $M$ (these lines are inferred from Fig. \ref{fig:CoolChain}). Fig. \ref{fig:obs} demonstrates that a substantial fraction of magnetic white dwarfs (in a volume-complete sample) are actually younger than currently thought by about a Gyr (or, equivalently, by about 10 per cent). 

\section{Conclusions}\label{sec:summary}

During the first few Gyr of white dwarf cooling, heat flows outwards from the conducting degenerate core, through a radiative bottleneck, to a convective envelope which gradually expands inward. Eventually, the convective envelope directly couples to the degenerate core -- eliminating the radiative bottleneck \citep{Fontaine2001}. This convective coupling between the surface and the core moderates the decrease in the white dwarf's luminosity over time $L(t)$, effectively delaying its observed cooling by $\Delta t\sim\textrm{Gyr}$ (though the drop in the central temperature actually accelerates).
Magnetic fields are known to inhibit convection, such that they may alter the cooling rate $L(t)$. However, due to the radiative bottleneck, even the strongest measured white dwarf magnetic fields have no effect until convective coupling occurs \citep{Tremblay2015}.

\citet{Tremblay2015} estimated the magnetic field strength $B$ required to disrupt convection and change the cooling time (after convective coupling) by comparing the magnetic energy density to the pressure $B^2\sim 8\upi P$ at edge of the degenerate core, where the electron degeneracy parameter $\eta\approx 0$. Here, on the other hand, we considered the \citet{GoughTayler1966} extension for Schwarzschild’s convective instability criterion in the presence of magnetic fields. By implementing this criterion in the \textsc{mesa} stellar evolution code, we found that even much weaker fields $B^2\ll 8\upi P$ may open a substantial radiative window that decouples the degenerate core from the convective envelope. Specifically, old white dwarfs with magnetic fields $B\gtrsim 1\textrm{ MG}$ cool down similarly to fully radiative stars, and are therefore younger by about a Gyr than non-magnetic white dwarfs with the same luminosity. 

\citet{BagnuloLandstreet2022} showed that the frequency and strength of magnetic fields increase with age, such that old white dwarfs with fields exceeding this critical value are common \citep[see also][]{Berdyugin2023,Berdyugin2024}. Specifically, we identified several white dwarfs in their volume-limited sample that are younger by $0.5-1\textrm{ Gyr}$ than currently thought due to magnetic inhibition of convection around their degenerate cores.
This difference -- about 10 per cent of the total cooling time -- is comparable to other effects that are currently being studied, such as the cooling delay due to carbon--oxygen phase separation during the crystallization of the core \citep[e.g.][]{Bauer2023}. Incorporating the magnetic inhibition of convective energy transfer consistently in white dwarf cooling models is therefore crucial for accurate cosmic chronology \citep{Winget1987,HansenLiebert2003,Hansen2004}, as well as for determining the origin of the magnetic fields \citep[e.g. fig. 2 in][]{BlatmanGinzburg2024a}. Accounting for this effect is also essential when comparing theory to empirical tests of white dwarf cooling \citep{Barrientos2024}.
We emphasize that even with magnetic fields treated properly, reliably determining the mass, temperature, and age of white dwarfs with low effective temperatures $T_{\rm eff}\lesssim 5,000\textrm{ K}$ remains challenging \citep[e.g.][]{O'Brien2024}.

Throughout this letter we were agnostic to the origin of the magnetic field. However, several promising mechanisms rely on compositional (as opposed to thermal) convection to generate the magnetic field through a dynamo \citep{Isern2017,Fuentes2023,Fuentes2024,Castro-Tapia2024,Lanza2024}. This compositional convection is confined to much deeper layers of the star \citep{Isern2017,BlatmanGinzburg2024a,BlatmanGinzburg2024b}, and is therefore decoupled from the atmospheric thermal convection discussed here. Magnetic fields that are much stronger than currently observed $B>10^9\textrm{ G}$ may have additional effects on white dwarf cooling \citep{Bhattacharya2018}; such strong fields are beyond the scope of this letter.

\section*{Acknowledgements}

We thank the organizers and participants of the `Current challenges in the physics of white dwarf stars' workshop in Santa Fe (March 2024) for an excellent meeting which inspired this work. We also thank the reviewer, Pier-Emmanuel Tremblay, for pointing out several important issues. We acknowledge support from the Israel Ministry of Innovation, Science, and Technology (grant No. 1001572596), and from the United States -- Israel Binational Science Foundation (BSF; grant No. 2022175).

\section*{Data Availability}

The data underlying this article will be shared on reasonable request to the corresponding author.



\bibliographystyle{mnras}
\input{OldMag.bbl}





\bsp	
\label{lastpage}
\end{document}